\documentclass[pra,twocolumn,showpacs,superscriptaddress,amsfonts,amsmath,floatfix]{revtex4-1}

\usepackage{graphicx}

\begin{document}

\title{Nonadiabatic creation of macroscopic superpositions with strongly correlated 1D bosons on a ring trap}
\author{C. Schenke}
\affiliation{Universit\'e Grenoble 1/CNRS, LPMMC, UMR 5493, B.P.~166, 38042 Grenoble, France}
\author{A. Minguzzi}
\email{anna.minguzzi@grenoble.cnrs.fr}
\affiliation{Universit\'e Grenoble 1/CNRS, LPMMC, UMR 5493, B.P.~166, 38042 Grenoble, France}
\author{F.W.J. Hekking}
\affiliation{Universit\'e Grenoble 1/CNRS, LPMMC, UMR 5493, B.P.~166, 38042 Grenoble, France}

\begin{abstract}
We consider a strongly interacting quasi-one dimensional Bose gas on a tight ring trap
subjected to a localized barrier potential.  We explore the possibility to form  a
macroscopic superposition of a rotating and a nonrotating state under 
nonequilibrium conditions, achieved by a sudden quench of the barrier velocity. Using an
exact solution for the dynamical evolution in the impenetrable-boson (Tonks-Girardeau)
limit, we find  an expression for the many-body wavefunction
corresponding to a superposition state.  The superposition  is formed when
the  barrier velocity is tuned close to multiples of integer or half-integer number of
Coriolis flux quanta. As a consequence of the strong interactions, we find that (i)  the
state of the system can be mapped onto  a macroscopic superposition of two Fermi spheres,
rather than two macroscopically occupied single-particle states as in  a weakly
interacting gas, and (ii) the barrier velocity should be larger than the sound velocity
to better discriminate the two components of the superposition.
\end{abstract}
\maketitle

\section{Introduction}
Macroscopic superpositions are at the heart of quantum information devices as they
realize quantum two-level systems (Qubits). Qubits can either be of single particle
nature (single atom, single spin), or realized using collective degrees of freedom whose
low-energy spectrum reduces to two discrete states (eg collective internal state
transitions for Rydberg atoms  \cite{BriMolSaf07}, and current states in superconducting
SQUIDs \cite{Orl99}).  Collective-mode superpositions are expected to be better protected
against some forms of decoherence such as particle losses, since quantum correlations are
spread over several single particle modes and the  loss of one particle does not imply
the destruction of the  collective mode, hence allowing the superposition to survive. On
the other hand, multimode superpositions imply the use of several different single
particle states, and hence typically have  a limited degree of entanglement.

Ultracold atomic gases are  interesting candidates for the realization of macroscopic superpositions due to
their high purity and tunability. Most of the current proposals are based on two-mode Bose-Josephson junctions
\cite{Mic03,PiaPezSme08,Ferrini08}. Experimental advances in the realization of ring traps \cite{SauBarCha01,Gup05,ArnGarRii06,Mor06,Paint,Ram11,She11} make it realistic to consider other macroscopic superpositions, eg the (collective-mode)  superposition of superflow states carrying different values of angular momentum \cite{HalBurDun06,SolMoz10a,SolMoz10b,HalErnBra10}, where the coupling between  angular-momentum states is provided by a localized barrier which breaks translational invariance; an artificial gauge field (or rotation) \cite{DalGer10} gives rise to  tunability  equivalent to  magnetic flux in a SQUID.
As a consequence of the ring periodicity, the energy levels of the many-particle system as a function of the
flux $\Phi$  associated with the artificial gauge field are periodic with period $\Phi_0=2 \pi \hbar /m$, $m$
being the atomic mass.
In an adiabatic protocol, equally-weighted superposition states can  be realized  by  tuning the flux near a
half-integer value of the ratio $\Phi/\Phi_0$, in correspondence to an avoided level-crossing~\cite{HalErnBra10}.

In the case of a quasi-1D tightly-confining ring trap the possibility of creating
macroscopic superpositions has been considered in detail, both for lattice and continuum
models \cite{HalBurDun07}. With respect to an ideal Bose gas,  weak repulsive
interactions induce small energy-level splittings which are harmful to the
superpositions. Quite interestingly, in the strongly-interacting limit of impenetrable
bosons (or Tonks-Girardeau limit) this drawback is overcome.  Due to its fermionized
character \cite{Gir1960}, the Tonks-Girardeau (TG)  gas displays the same energy
 splittings as for noninteracting bosons \cite{HalErnBra10}. Moreover, due to
its impenetrability, two- and three-body losses are suppressed in a TG gas
\cite{GanShlyap03}, thus eliminating one of the main sources of decoherence in ultracold
gases. The TG gas is therefore a very promising candidate for the realization of
macroscopic superpositions of current states, as can be obtained, eg, by setting into
motion a localized barrier potential.  In the TG limit, creation of stationary
superposition states with velocity $0$ and $2\pi \hbar /mL$, $L$ being the ring
circumference    can be obtained by an adiabatic switching on of the barrier to a
velocity $\pi \hbar /mL$ corresponding to half a Coriolis flux quantum
\cite{HalErnBra10}. Such superpositions have maximal useful correlations for
interferometry \cite{CooHalDun10,CooHalBra11}, with applications to ultra-precise  atomic
gyroscopes.

We focus in this work  on a  sudden switch  on of the barrier motion to velocity $v$
multiple of $\pi \hbar /mL$. We find that it  gives rise to  Rabi-like oscillations
between states  with velocity components $0$ and  $2v$, similarly to what has been
observed for a superlattice ring \cite{NunReyBur08}. At  specific times, we find an
equally-weighted  macroscopic superposition of two multiparticle states with velocity
components $0$ and  $2v$. In view of the multimode aspects of the strongly correlated TG
state, several questions are  open regarding such a novel superposition, in particular on
its nature and on its degree of entanglement. For example, similarly
to a Fermi gas, impenetrable bosons cannot occupy the same single particle state, hence, different from  Ref.\cite{NunReyBur08}, the macroscopic
superposition is not expected to be close to the usual ``NOON'' state,
$|NOON\rangle\propto [(b^\dagger_0)^N+(b^\dagger_{q_0})^N]|vac\rangle$  where all the $N$
atoms occupy the $q=0$ or  the $q_0=2 m v/\hbar$ single particle level, $b^\dagger_q$
being the creation operator of a boson with momentum $q$.

These questions can be addressed by our  fully microscopic,
analytical solution for the dynamical evolution by mapping onto a Fermi gas
\cite{GirWri00}.  We obtain the exact expression for the wavefunction of the
superposition state, which schematically reads $|\Psi\rangle\propto \tilde {\cal A}
\Pi_{-k_F<k<k_F}[c^\dagger_k+c^\dagger_{k+q_0}+c^\dagger_{-k+q_0}]|vac\rangle$, where
$c^\dagger_k$ is the creation operator of a fermion with momentum $k$, $ \tilde {\cal A}$
is the mapping function from fermions to bosons  and $k_F=\pi N/L$ is the Fermi
wavevector of the mapped Fermi gas. Such  an entangled, correlated many-body
superposition is only accessible through an out-of-equilibrium drive. Furthermore, our
microscopic approach allows to obtain the time of formation of the superposition, and to
set constraints on the excitation process. Finally, it allows to  simulate the
time-of-flight signal which is the standard probe used in  experiments.

\section{Exact dynamical solution for nonadiabatic stirring}
We consider $N$ impenetrable bosons on a ring  of circumference $L$ at zero temperature, subjected to the stirring delta-barrier potential $U(x,t)=U_0 \delta(x-vt)$. The Hamiltonian is
\begin{equation}
\hat{H}_B=\sum_{j=1}^{N}\left[-\frac{\hbar^2}{2m}\frac{\partial^2}{\partial x_{j}^2}
+U(x_{j},t)\right]+\sum_{j<\ell}g \delta (x_j-x_\ell)
\end{equation}
 where in particular the Tonks-Girardeau regime corresponds to the limit $g\to\infty$ and the interaction term can be replaced by the condition that the many-body wavefunction vanishes at contact between each pair of bosons, $\Psi_B(...x_j=x_\ell...)=0 $. As we want to describe the ring geometry, we impose periodic boundary conditions ie $\Psi_B(...x_j...)=\Psi_B(...x_j+L...)$ for any $j=1...N$. The exact solution for the many-body wavefunction is obtained by the time-dependent Bose-Fermi mapping \cite{GirWri00},
\begin{equation}
\label{psi_manybody}
\Psi_B(x_1,...x_N,t)= {\cal A} (1/\sqrt{N!})\det[\psi_l(x_m,t)]
\end{equation}
where ${\cal A}=\Pi_{j<\ell}{\rm sgn}(x_j-x_\ell)$ is the mapping function, with ${\rm sgn}(x)=1$ for $x>0$ and ${\rm sgn}(x)=-1$ for $x<0$. The orbitals $ \psi_l(x,t)$ are obtained from the solution of the time-dependent Schroedinger equation
\begin{equation}
\label{tdstirring}
i\hbar \partial_t \psi_l(x,t)=\left(-\frac{\hbar^2}{2m}  \partial_x^2 + U_0 \delta(x-vt)\right) \psi_l(x,t).
\end{equation}
We choose as initial condition for the TG gas its ground state in the presence of a nonmoving barrier, ie a Fermi sphere for the mapped Fermi gas built with the orbitals  $\psi_l(x,0)=\phi^{(0)}_l(x)$,  eigenvectors of the Schroedinger equation for a  nonmoving barrier with eigenvalues $E_l^{(0)}=\hbar^2 {k_l^{(0)}}^2/2m$, for $l=1...N$. The barrier is then suddenly set into motion at time $t=0^+$.
The use of two unitary transformations  ${\cal U}_1=e^{-i  p v t/\hbar}$, ${\cal U}_2=e^{i  m v x/\hbar}$ maps the problem onto a stationary one with twisted boundary conditions, enabling to express the solution of Eq.(\ref{tdstirring}) as
\begin{equation}
\label{eq:orbitals}
\psi_l(x,t)=e^{i q x} e^{-i q^2 t/2m}\sum_j c_{jl}e^{-i E_j t} \phi_j(x-vt),
\end{equation}
where we have defined the quasimomentum $q=mv/\hbar$. The orbitals $  \phi_j(x)$ are the solutions of
$E_j  \phi_j(x)=(-(\hbar^2/2m)  \partial_x^2 + U_0 \delta(x))  \phi_j(x)$  with twisted boundary conditions $  \phi_j(x+L)=e^{-iqL}   \phi_j(x)$, and read
\begin{equation}
\label{eq:phij}
\phi_j(x)=\frac{1}{{\cal N}_j}
\begin{cases}
e^{iq\frac{L}{2}}(e^{i(k_j(x+\frac{L}{2})}+A_{j} e^{-i k_j(x+\frac{L}{2})}) \!\!\!\!\!\!
&\mbox{in}\,  [-\frac{L}{2},0) \\
e^{-iq\frac{L}{2}}(e^{i(k_j(x-\frac{L}{2})}+A_{j} e^{-i k_j(x-\frac{L}{2})}) \!\!\!\! & \mbox{in}\,  [0,\frac{L}{2}], \end{cases}
\end{equation}
with normalizations $ {\cal N}_j=\sqrt{L(1+ A_j^2+ 2 A_j \frac{\sin(k_jL)}{k_jL})}$, amplitudes $A_j=\sin[(k_j+q)L/2]/\sin[(k_j-q)L/2]$, and wavevectors $k_j$ given by the solution of the transcendental equation
$k_j=\lambda  \sin(k_jL)/(\cos(qL)-\cos(k_jL))$, which determines the  energy eigenvalues  $ E_j=\hbar^2 k_j^2/2m$, with $\lambda= mU_0/\hbar^2$. Information about the initial condition enters Eq.(\ref{eq:orbitals}) through the time independent overlaps  $c_{jl}=\langle  \phi_j|e^{-iqx}|\phi_l^{(0)}\rangle$.

\begin{figure}
\centerline{\includegraphics[width=0.4\textwidth]{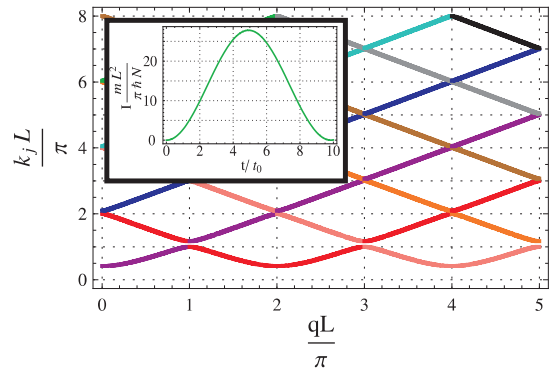}}
\caption{(Color online) Single-particle excitation spectrum $k_j L/\pi$ versus stirring momentum $qL/\pi$. The color code denotes the branches of given angular momentum in the zero barrier limit. The inset shows the short-time evolution (time in units of $t_0=mL^2/\pi\hbar$) of the integrated  particle current (in units of $\pi \hbar N/mL^2$).}
\label{fig1}
\end{figure}

\section{State of the system under the stirring drive}
We first consider a velocity close to a special value  $v=\hbar n \pi/ mL$ with integer $n$, which corresponds to half integer values of the Coriolis flux ratio $\Phi/\Phi_0$, with $\Phi=vL$. We focus on the  small barrier limit $\lambda L \lesssim 1$. As is seen in Fig.\ref{fig1}, this choice of $v$ corresponds to several avoided level crossings of  single particle states.
 As a consequence of the sudden quench of the barrier velocity, we find that the TG gas, which initially occupies the zero-momentum Fermi sphere  of the mapped Fermi system, oscillates between  two $N$-particle Fermi spheres, one centered at $k=0$ and the other at $k=2 q$, realizing at half oscillation an equally weighted  macroscopic superposition of the two Fermi spheres.

The derivation reads as follows. According to the expression for the overlaps  $c_{jl}$, the states excited  under the effect of the stirring drive are  fixed by quasimomentum conservation  $k_j=  k_l^{(0)} \pm q$ for $v<v_F$, or  $k_j= \pm k_l^{(0)}+q$ for $v>v_F=N\hbar\pi/mL$.  In detail, taking for simplicity $v>v_F$ which will turn out to be the most favourable situation, we find that to leading order in $\lambda L$  only four states $j$  are coupled to each single-particle level $l$ of the initial-state Fermi sphere, with coefficients  $|c_{jl}|=1/2$ for $j=n\pm 2 {\rm Int}[l/2]$ and $j=n+1\pm 2 {\rm Int[l/2]}$, Int$[..]$ denoting the integer part; with the exception of the lowest  state $l=1$, where   $|c_{jl}|=1/\sqrt{2}$ for $j=n$ and  $j=n+1$. For each level $k_j$ we know its  momentum (hence angular momentum) components from the analysis of the zero-barrier limit, where a true level crossing occurs of two states of well defined angular momentum. For example, for $v=4\pi \hbar/mL$ and $N=3$, the level $l=1$ is coupled by the stirring barrier to  the states with $j=4$ and 5 which are both an equal-weight superposition of states with momentum $k=0$ and $k=8\pi/L$. Similarly,  the levels $l=2$ and 3 yield an equal-weight superposition of states with $k=2\pi/L$ and $6\pi/L$ for the lowest-energy  doublet and of $k=-2\pi/L$ and $10 \pi/L$ for the highest-energy  doublet. Summing up all the contributions, we find that each momentum state has the same occupation, and the momentum occupation distribution is a superposition of the two Fermi spheres $\{-2\pi/L,0, 2\pi/L\}$ and  $\{6\pi/L,8\pi/L, 10\pi/L\}$. A similar reasoning holds for arbitrary barrier velocities and particle numbers (chosen odd to ensure proper boundary conditions on the mapped Bose gas), leading  to the occupation of two   Fermi spheres centered at $0$ and $2 q$.  From Eqs.(\ref{psi_manybody}) and (\ref{eq:orbitals}), the many-body wavefunction of the superposition is finally obtained by mapping onto a Fermi gas where each atom occupies a different superposition of a few (typically four, in the weak barrier limit) single particle orbitals.

The detailed dynamics of the system is also simply described, according to the values of the overlaps $c_{jl}$,  in terms of the occupation of a few momentum states for each single-particle state $l$. Using Eq.~(\ref{eq:orbitals}), the time dependence eg of the particle current density $j(x,t)=(\hbar/m){\rm Im}\sum_l \psi^*_l(x,t) \partial_x \psi_l(x,t)$ is fixed by the time-evolution factors $e^{-i(E_j-E_{j'})t/\hbar}$. The typical time scale is fixed by the  energy level splitting associated to the highest two levels occupied  through the stirring drive. The short-time behaviour of the integrated current $I(t)=(1/L) \int dx j(x,t)$ is illustrated in the inset of Fig.\ref{fig1}; the time evolution is not purely sinusoidal due to the multimode nature of the superposition.

Consider now the off-resonant case $\hbar (n-1) \pi/ mL<v< \hbar n \pi/ mL$. In this case a weak barrier  does not transfer angular momentum to the gas and  the momentum occupation distribution is  a single Fermi sphere centered at $k=0$. This is readily derived by inspecting the overlaps $c_{jl}$, yielding $|c_{j,1}|=1$ for $j=n$, and $|c_{jl}|=1/\sqrt{2}$ for $l>1$ with $j=n \pm 2 {\rm Int} [l/2]$.

We note that the state of the system found under a sudden switching on
of the barrier velocity is very different  from the one obtained by an adiabatic turning
on, where the lowest $N$ single particle energy levels are populated and   the momentum
occupation distribution is  either a single Fermi sphere centered at $k=q$ for even $n$,
or  a superposition of two Fermi spheres centered at $k=q-\hbar \pi/L$ and $k=q+\hbar
\pi/L$ for odd $n$.

As a partial summary, we find that the creation of macroscopic superpositions is
efficient only in the vicinity of $v\simeq  \hbar n \pi /mL$. Fluctuations on the barrier
velocity might degrade the quality of the superposition, leading in particular to a
different weight for  the two components of the superposition. The velocity window useful
for the excitation depends on the details of the barrier, eg on  the barrier height. We
find that  an  increase of the barrier height improves the robustness of
the superposition state with respect to barrier velocity fluctuations.

\begin{figure}
\includegraphics[width=0.20\textwidth]{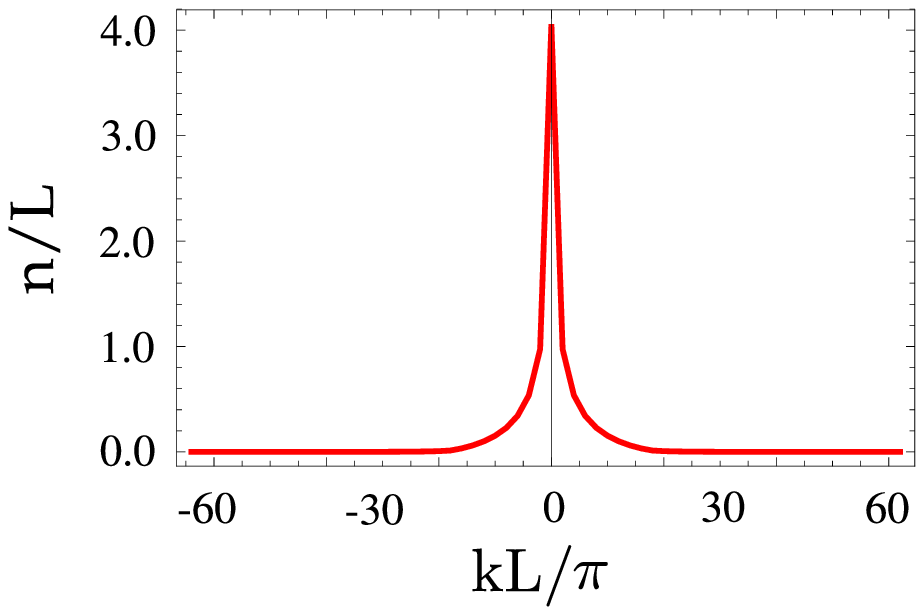}\includegraphics[width=0.25\textwidth]{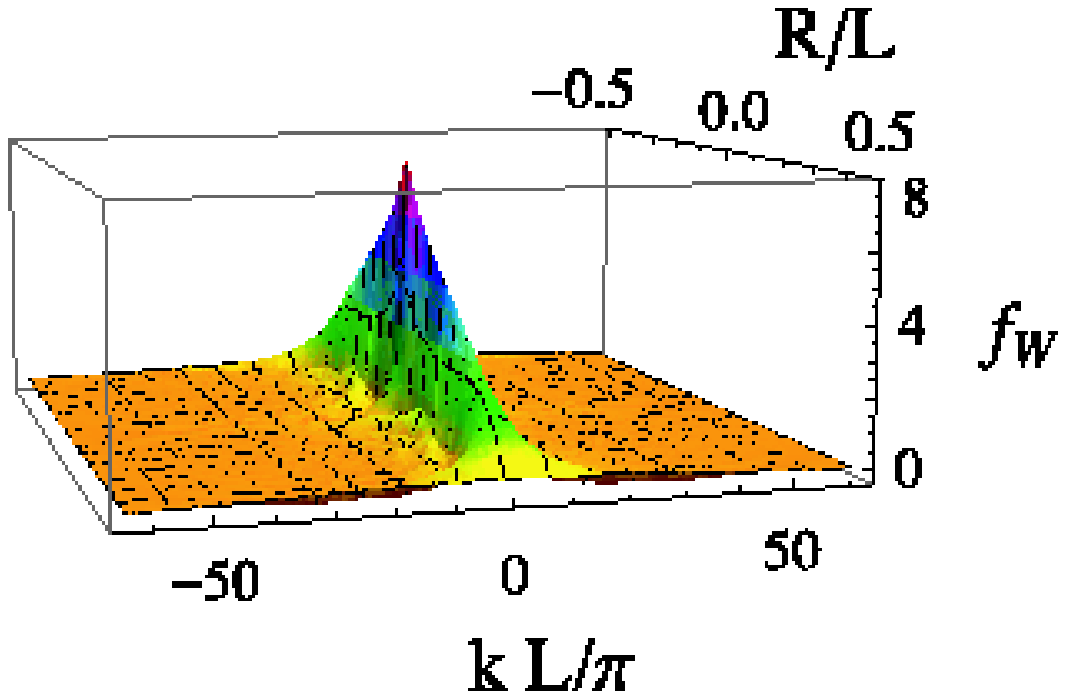}
\includegraphics[width=0.20\textwidth]{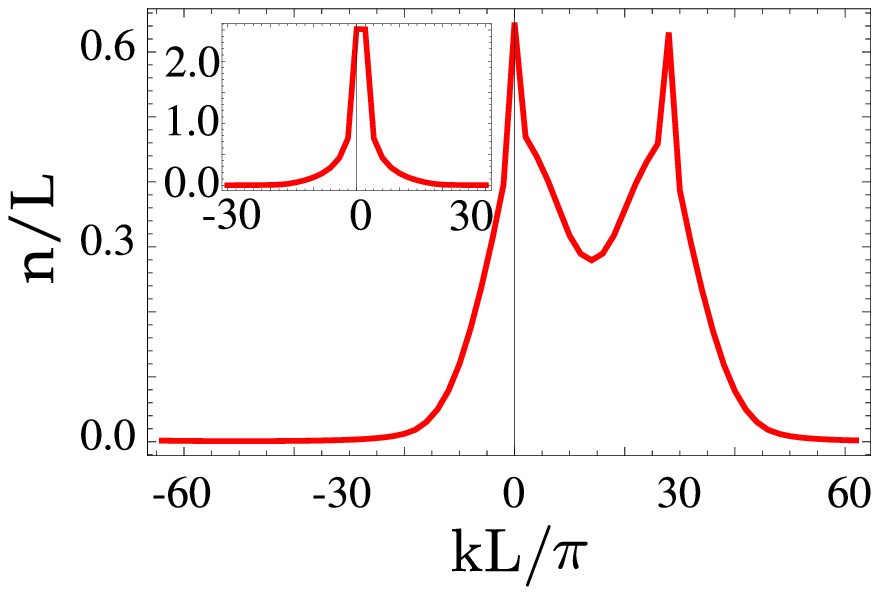}\includegraphics[width=0.25\textwidth]{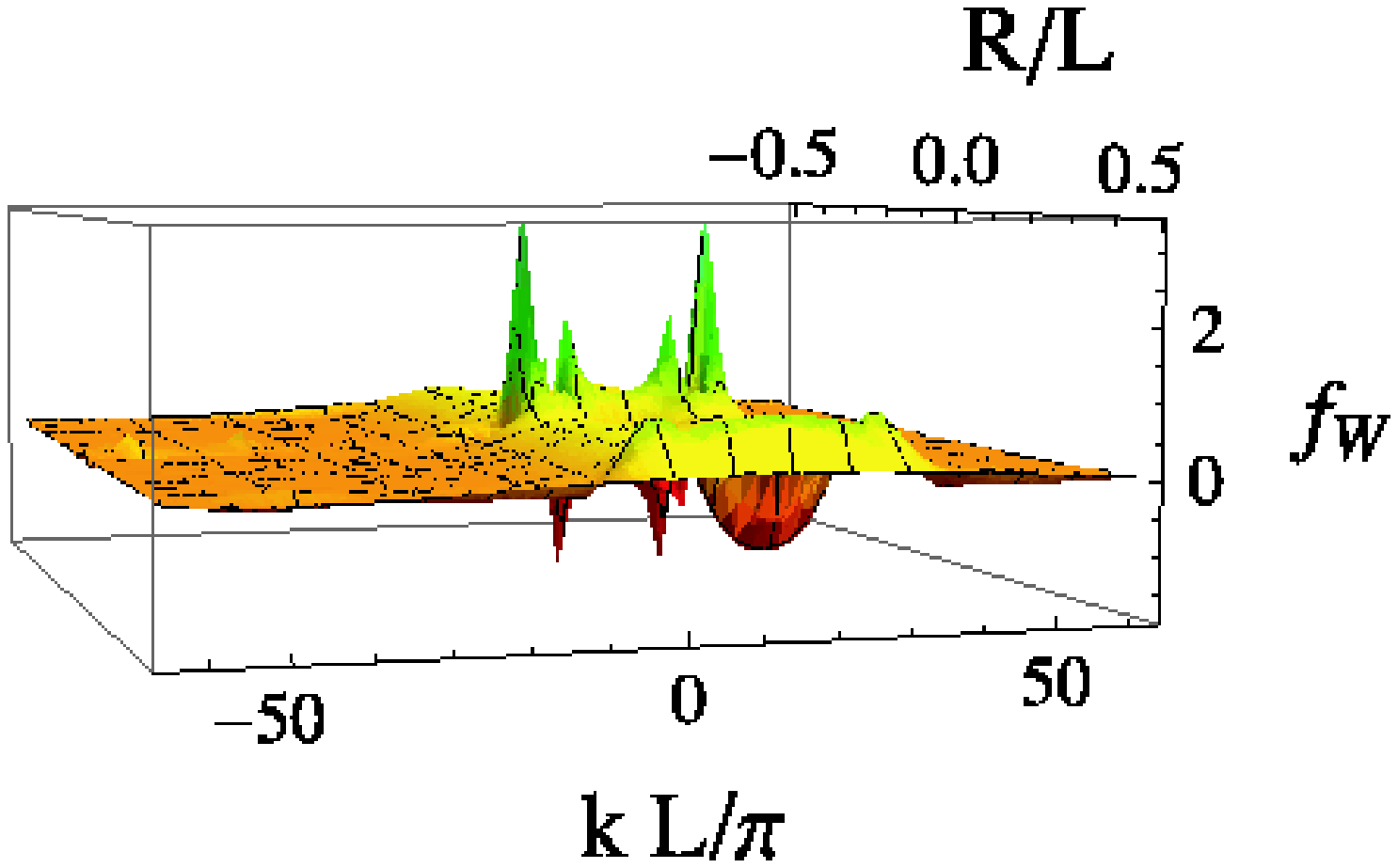}
\includegraphics[width=0.20\textwidth]{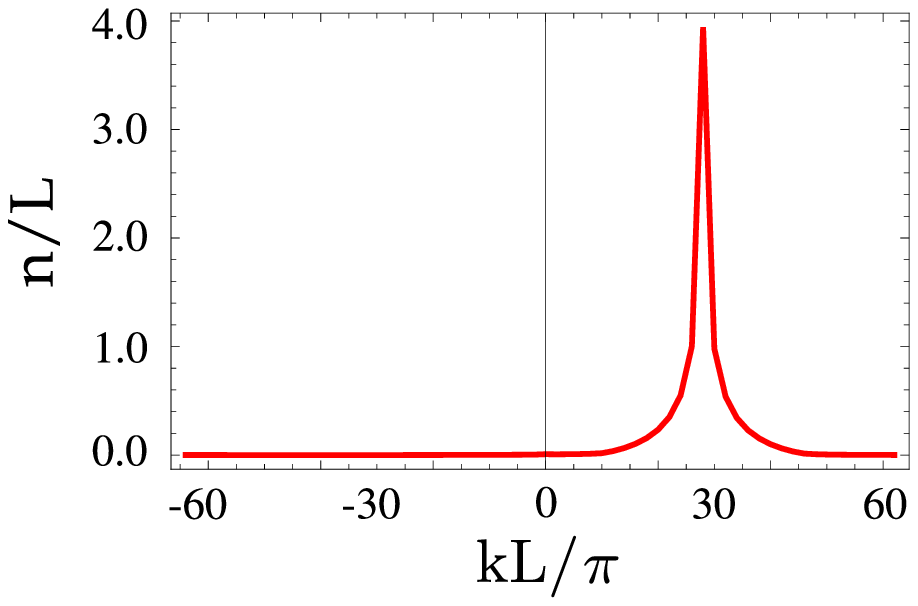}\includegraphics[width=0.25\textwidth]{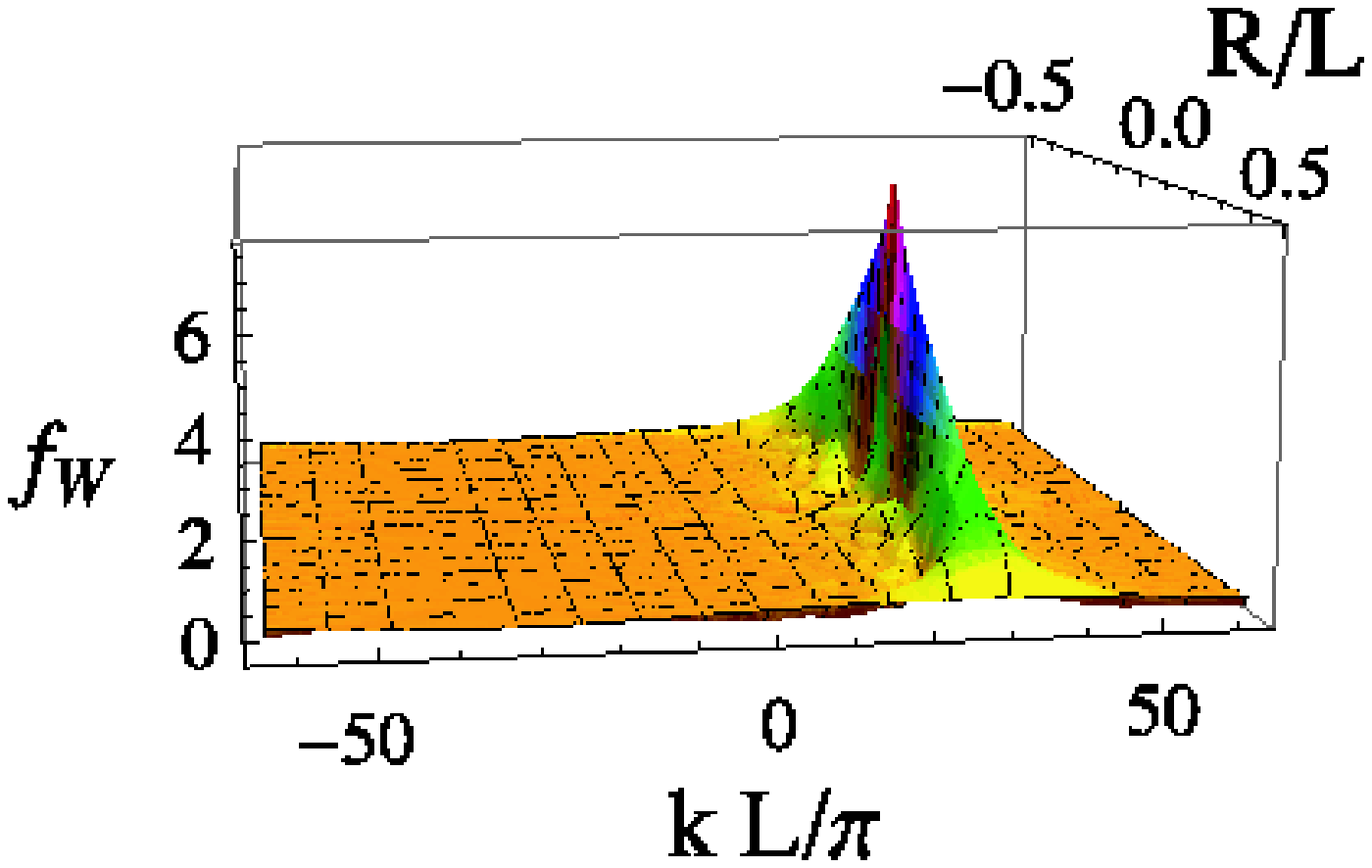}
\caption{(Color online) Time-dependent momentum distribution (left panels, in units of $L$), as a function of the wavevector $kL/\pi$, and corresponding Wigner function (right panels, dimensionless) as a function of $X/L$ and $kL/\pi$ at times $t/t_0=0$, 2.46 and 4.92 for N=9 particles and stirring velocity $v=14 \hbar \pi/mL$. The inset of the third panel shows the momentum distribution at the time of an equally-weighted superposition for stirring velocity $v=\hbar \pi/mL$}
\label{fig2}
\end{figure}

\section{One-dimensional momentum distribution and  Wigner function}
We illustrate the formation of the macroscopic superposition by following the dynamical evolution of the 1D momentum distribution, $n(k,t)=\int dx \int dy e^{i k(x-y)} \rho_1(x,y,t)$. It is defined in terms of the (time-dependent) one-body density matrix $\rho_1(x,y,t)$, which can be efficiently calculated according to  \cite{PezBul07}
\begin{equation}
\rho_1(x,y,t)=\sum_{l,l'=1}^N \psi_l^*(x,t) A_{l,l'}(x,y,t) \psi_{l'}(y,t),
\end{equation}
with
$A_{l,l'}(x,y,t)=[\det P] (P^{-1})^T_{l,l'}$ and $P_{l,l'}(x,y,t)=\delta_{l,l'}-2 \int_x^y dx'\psi_l^*(x',t)  \psi_{l'}(x',t)$.  As shown  in Fig.\ref{fig2}, during the time evolution induced by a stirring velocity close to an integer multiple of $\hbar \pi/ mL$ the momentum distribution evolves from a single peak  at $k=0$ to a single peak at $k=2q$, displaying at intermediate times a double peak structure, reflecting the superposition of the two Fermi spheres of the mapped Fermi gas. The peaks in the momentum distribution, associated with the bosonic nature of the gas,  allow to well identify the two components.  However, since the width of the TG momentum distribution is  the same as the fermionic one, in order to better resolve the superposition one needs stirring velocities larger than twice the Fermi velocity ie the sound velocity of the TG gas (see the inset in Fig.\ref{fig2} for an exemple of momentum distribution for a superposition state in the case of a small velocity $v=\pi\hbar/mL$).

The observation of a double peak structure in the momentum distribution does not necessarily imply the existence of quantum correlations between the two Fermi spheres \cite{note}. In order to  evidence the nonclassical nature of the macroscopic superposition we compute its Wigner function,
\begin{equation}
f_W(X,k,t)=\int dr  e^{i kr} \rho_1(X+r/2,X-r/2,t).
\end{equation}
At the time of the equally-weighted superposition, the Wigner function displays some negative regions, see again Fig.\ref{fig2}. This illustrates the  quantum correlations between the two Fermi spheres, which could be quantified eg following~\cite{LeeJeo11}.

\begin{figure}
\includegraphics[height=0.25\textwidth,width=0.35\textwidth]{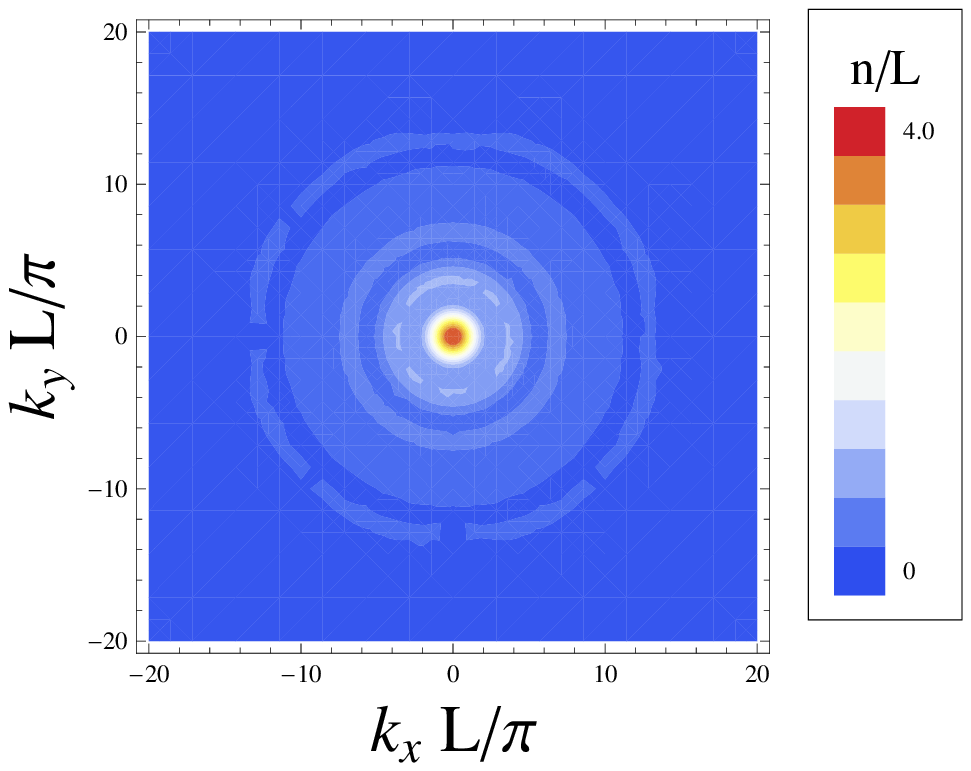}
\includegraphics[height=0.25\textwidth,width=0.35\textwidth]{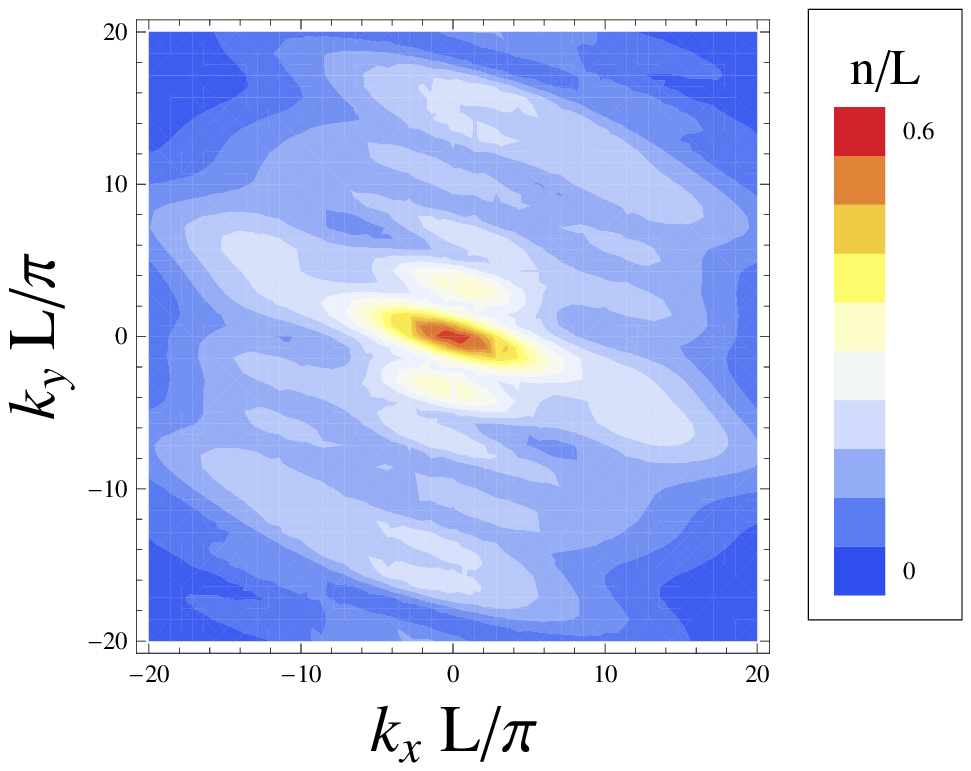}
\includegraphics[height=0.25\textwidth,width=0.35\textwidth]{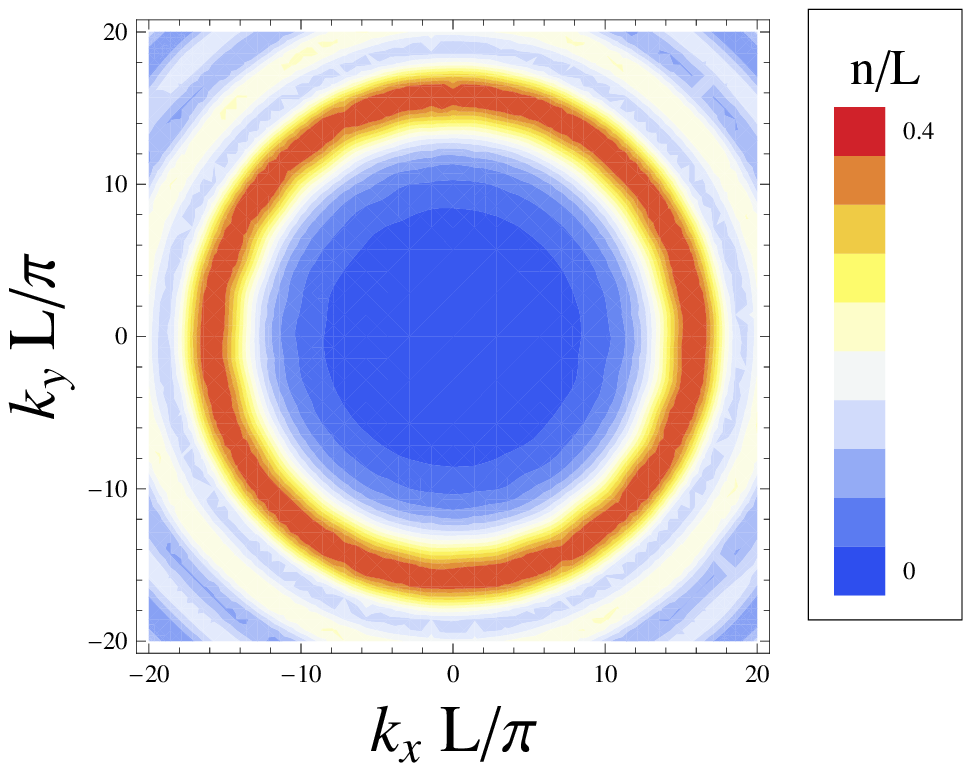}
\caption{(Color online) TOF images in the $k_x$ $k_y$ plane (in units of $\pi/L)$ for $N=9$ bosons stirred at velocity $v=14\pi\hbar /mL$, at times $t/t_0=0$, 2.46, 4.92, from top to bottom.}
\label{fig3}
\end{figure}

\section{Observability in ultracold atomic gases}
Current experimental detection techniques are based on  time-of-flight (TOF)  images, obtained by releasing the confining potential.  Neglecting the effects of interactions after the release from the trap, we describe the spatial distribution of the atomic cloud  after expansion as the momentum distribution of the gas before expansion. In the case of the ring we have
\begin{equation}
n_{TOF}({\mathbf k})=\int d^3x \int d^3x'  e^{i {\mathbf k}\cdot ({\mathbf x}-{\mathbf x'})} \rho_1^{ring}({\mathbf x},{\mathbf x'},t)
\end{equation}
where the expression in cylindrical coordinates of the 3D one-body density matrix on a tight ring trap of radius $R$ is  $\rho_1^{ring}(r,\theta,z;r',\theta',z';t)=\delta(r-R)\delta(r'-R)\delta (z) \delta (z') \rho_1(R\theta,R\theta',t)$. Figure \ref{fig3} shows the TOF images corresponding to various stirring times, illustrating  the transition between a  zero-current state at initial time to a state of angular momentum ${\cal L}/N=2m v R$. The initial   peak at ${\mathbf k}=0$ deforms spirally and finally tends to a ring, the latter in agreement with the predictions of \cite{SolMoz10a} for a state with well-defined current. Note that the TOF image of the equal-weight macroscopic superposition, represented in
the second panel in Fig.\ref{fig3}, is not simply obtained as a  combination of the TOF images of well-defined current  states (first and last panel in  Fig.\ref{fig3}), due to interference between the zero-current state and the state at velocity $2v$.

\section{Summary and perspectives} We have studied the nonadiabatic excitation of strongly interacting bosons on a ring by a sudden set into motion of a localized barrier potential at velocity $v$. If the velocity is suitably chosen, the state of the system oscillates between a zero-current state and a state of velocity $2v$, displaying a  superposition of the two states at intermediate times. Due to the strong interactions, the nature of the superposition state is very different from the one of a weakly interacting gas;  we find a superposition of two Fermi spheres rather the superposition of two macroscopically occupied single-particle levels. The superposition of current states is evidenced by a double-peak structure in the momentum distribution. Due to the underlying multimode nature of the state, the two peaks are resolved at best for stirring velocities larger than twice the sound velocity. We have also verified  the nonclassical nature of the superposition by the study of the Wigner function. Our results confirm the TG gas as a promising candidate  for applications to quantum-limited metrology and seem accessible to state of the art experiments on ultracold gases.
In perspective, it will be important to develop detecting techniques capable to estimate the relative weight of  the two components of the superposition. A measure of their coherence could be inferred by extending the full-counting method proposed in  \cite{Ferrini09}.

\begin{acknowledgments}
We acknowledge discussions with Edouard Boulat, Roberta Citro, Leonid Glazman, Vincent Lorent and Helene Perrin. We thank the PEPS-PTI project ``Quantum gases and condensed matter'', the MIDAS STREP project and the Handy-Q ERC project for financial support.
\end{acknowledgments}


\begin{thebibliography}{10}
\bibitem{note} An  analysis of the momentum distribution of  a statistical mixture of the same two states indeed shows some quantitative differences with respect to the momentum distribution of the macroscopic superposition. In particular, the momentum distribution of a statistical mixture  is lower in the region between the two peaks, while each peak is higher.
\end{thebibliography}
\end{document}